\documentclass[a4paper,11pt]{article}
\usepackage{style}
\usepackage{orcidlink}
\usepackage{graphicx}
\usepackage{placeins}
\usepackage{amsmath,amssymb}
\usepackage[backend=biber, style=numeric, sorting=none]{biblatex}
\DefineBibliographyStrings{english}{%
  in = {}
}
\renewbibmacro{in:}{}
\title{Probing new physics with dedicated data streams at CMS}

\author{Ali Eren SIMSEK\,\orcidlink{0000-0002-9074-2256}}
\onbehalf{for the CMS Collaboration}

\affiliation{Catholic University of America,\\
620 Michigan AVE NE, Washington, DC, United States of America}

\emailAdd{ali.eren.simsek@cern.ch}

\abstract{Signatures of new physics at the LHC are varied and, by nature, often very different from those of Standard Model processes. Novel experimental techniques, including dedicated data streams, are exploited to enhance the sensitivity of the CMS Experiment to search for such signatures. This report highlights the CMS results obtained using data collected at the LHC during Run-II and Run-III through the so-called ``Data Scouting'' and ``Data Parking'' strategies. These approaches have allowed us to set some of the strongest constraints to date for low-mass resonances in prompt and long-lived signatures.}

\FullConference{
Phenomenology 2025 Symposium (PHENO) \\
May 19 - 21, 2025 \\
University of Pittsburgh, PA, United States of America
}

\begin{document}
\maketitle

\section{Introduction}
\label{sec:intro}

The Compact Muon Solenoid (CMS) detector~\cite{CMS:Detector2008,CMS:DetectorRun3} at the CERN LHC records proton--proton collisions delivered at a bunch-crossing rate of about 40~MHz. Only a very small fraction of these collisions can be stored with the full detector readout and reconstructed offline, due to limitations in trigger bandwidth, data acquisition, and computing resources. Conventional trigger menus are therefore optimized for a broad physics program, but unavoidably lose sensitivity to signatures that are very common or that require low thresholds and large statistics, particularly in the low-mass regime. In other words, the normal data-taking stream is highly efficient for many analyses, but it necessarily sacrifices sensitivity to some new-physics scenarios in which interesting events are frequent but individually unremarkable from the trigger point of view. These constraints are enforced through the CMS trigger system, which consists of a hardware-based Level-1 (L1) trigger and a software-based high-level trigger (HLT)~\cite{Khachatryan_2017}. The L1 trigger, implemented in custom electronics, receives coarse-granularity information from the calorimeters and muon systems at the full LHC bunch-crossing rate and selects events at a maximum rate of about 100~kHz. The HLT then processes the corresponding full detector readout using algorithms similar to those employed offline and reduces the rate to approximately 1~kHz for permanent storage.

\begin{figure}[!htbp]
  \centering
  \includegraphics[width=\textwidth]{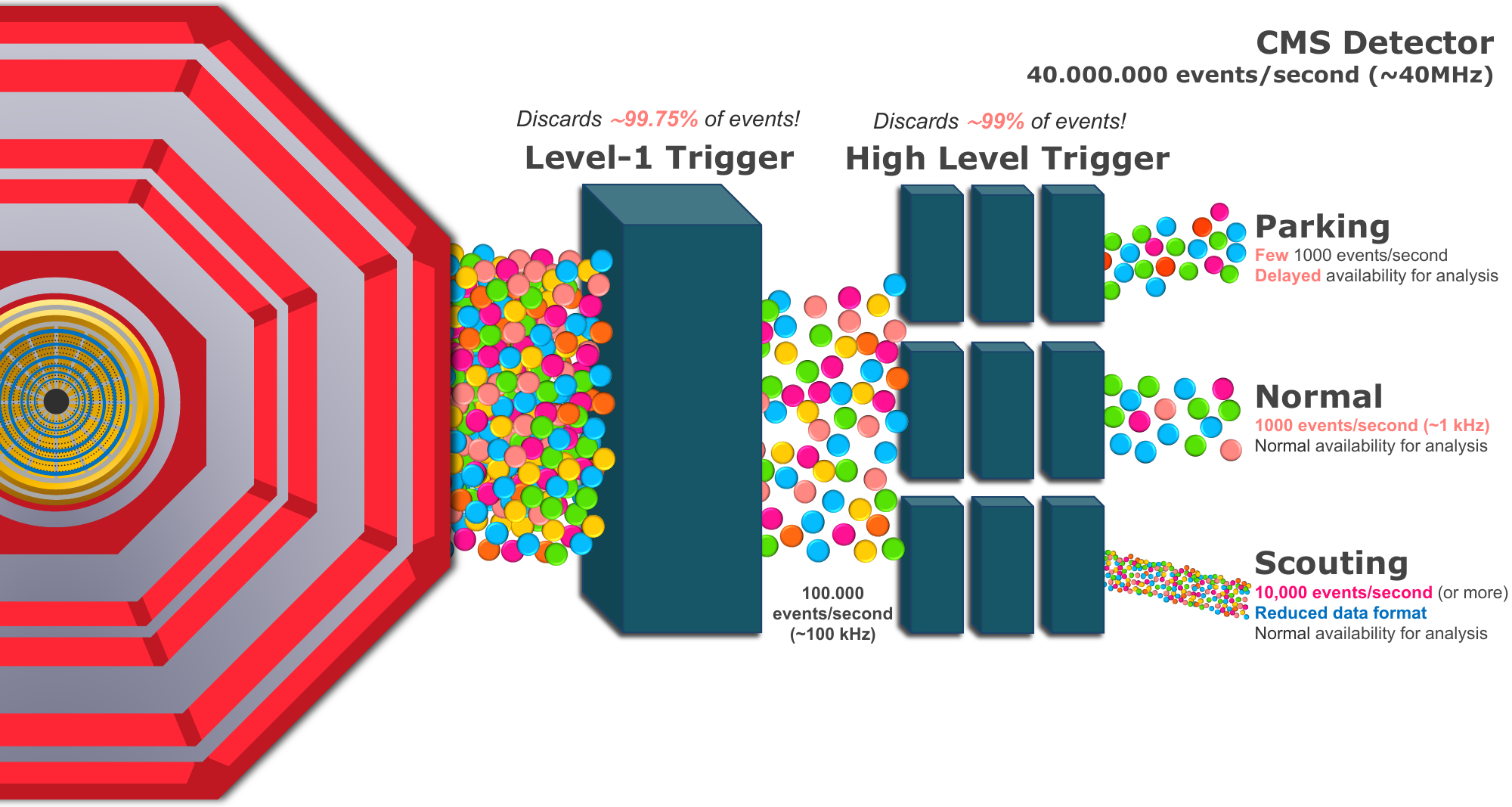}
  \caption{Schematic illustration of the CMS data flow, showing the L1 trigger, the high-level trigger, and the three main data streams: normal, parking, and scouting.}
  \label{fig:streams}
\end{figure}

To overcome this limitation of the trigger system, CMS has developed specialized data-taking strategies that exploit the flexibility of both the trigger and data-acquisition systems. The \emph{data scouting}~\cite{DataScouting_CMS} strategy records events with a highly reduced event content at much higher rates than the normal data stream. In data scouting, events are reconstructed online in the HLT and written to disk with a strongly reduced event content, typically including only the physics objects relevant for a given analysis and a minimal set of event-level information. By removing the full detector readout and detailed per-channel information, the average event size can be reduced by up to two orders of magnitude, enabling event rates of $\mathcal{O}(10~\text{kHz})$ without exceeding the available bandwidth. In practical terms, this means that scouting streams trade away detailed detector information in order to dramatically increase statistics for signatures that benefit from low thresholds and very large samples, such as low-mass resonances. The \emph{data parking} strategy is complementary to scouting and addresses the same underlying problem from a different angle~\cite{EXO-24-007_EnrichingCMSScouting_2025}. In parking streams, events are stored with the full raw detector readout, as in the normal stream, but are not reconstructed immediately. Instead, they remain in long-term storage until sufficient offline computing resources become available, at which point the data are reconstructed in bulk. This approach relaxes the CPU constraints from prompt reconstruction while preserving full detector information. It is particularly useful when lower trigger thresholds or additional specialized triggers are desired. Fig.~\ref{fig:streams} schematically illustrates the data flow from the detector through L1 and the HLT into these three data streams. The relative rates shown on the figure are representative of Run~2 and Run~3 configurations, and detailed information can be found in Ref.~\cite{EXO-24-007_EnrichingCMSScouting_2025}. 

Looking further ahead, the same philosophy is being extended in the CMS Phase-2 upgrade for the high-luminosity LHC. In addition to the normal, parking, and scouting streams, a Level-1 data-scouting system is being developed that streams selected L1 trigger objects at the full 40~MHz rate to a dedicated processing farm~\cite{40MHzScouting}. This stream is intended to provide very large samples at L1 resolution for detector monitoring, luminosity measurements, and targeted physics studies that would otherwise be limited by the L1 output rate.

\section{Physics results with dedicated data streams}
\label{sec:physics}

The physics impact of dedicated data streams can be illustrated with a set of representative searches for new phenomena that rely on either data scouting or data parking. These analyses focus on signatures that are especially sensitive to trigger and bandwidth limitations, such as low-mass resonances, high-rate final states, and long-lived particles that would otherwise be heavily suppressed or entirely missed by the standard trigger menu. By exploiting reduced event content, deferred reconstruction, and specialized trigger strategies tailored to these signatures, they open up regions of phase space that are difficult to access with the conventional CMS data stream and thereby substantially broaden the overall discovery potential of the experiment.

\subsection{Search for pair-produced multijet resonances}
\label{sec:paired_produced_multijet}

A search has been performed for pair-produced multijet resonances in proton--proton collisions at $\sqrt{s}=13$~TeV, using the full Run~2 data set corresponding to an integrated luminosity of 128~$\mathrm{fb}^{-1}$~\cite{EXO-21-004_MultiJetScouting_2024}. The analysis targets new colored particles that decay promptly to fully hadronic final states, motivated in particular by $R$-parity-violating supersymmetric scenarios with higgsinos, top squarks, and gluinos decaying to two or three quarks. To maximize sensitivity to low-mass resonances in the high-rate multijet environment, the search is based on the CMS data scouting stream, allowing much lower $H_{\mathrm{T}}$ thresholds and substantially higher event rates than in the normal stream. In this search, three final states are considered. Two of them use large-radius ($R=0.8$) jets and are optimized for ``merged'' topologies, where the hadronization products of two or three quarks are reconstructed as a single fat jet. The third channel is a ``resolved'' search for pair-produced three-jet resonances, using triplets of small-radius ($R=0.4$) jets. 

\begin{figure}[!htbp]
  \centering
  \includegraphics[width=0.32\textwidth]{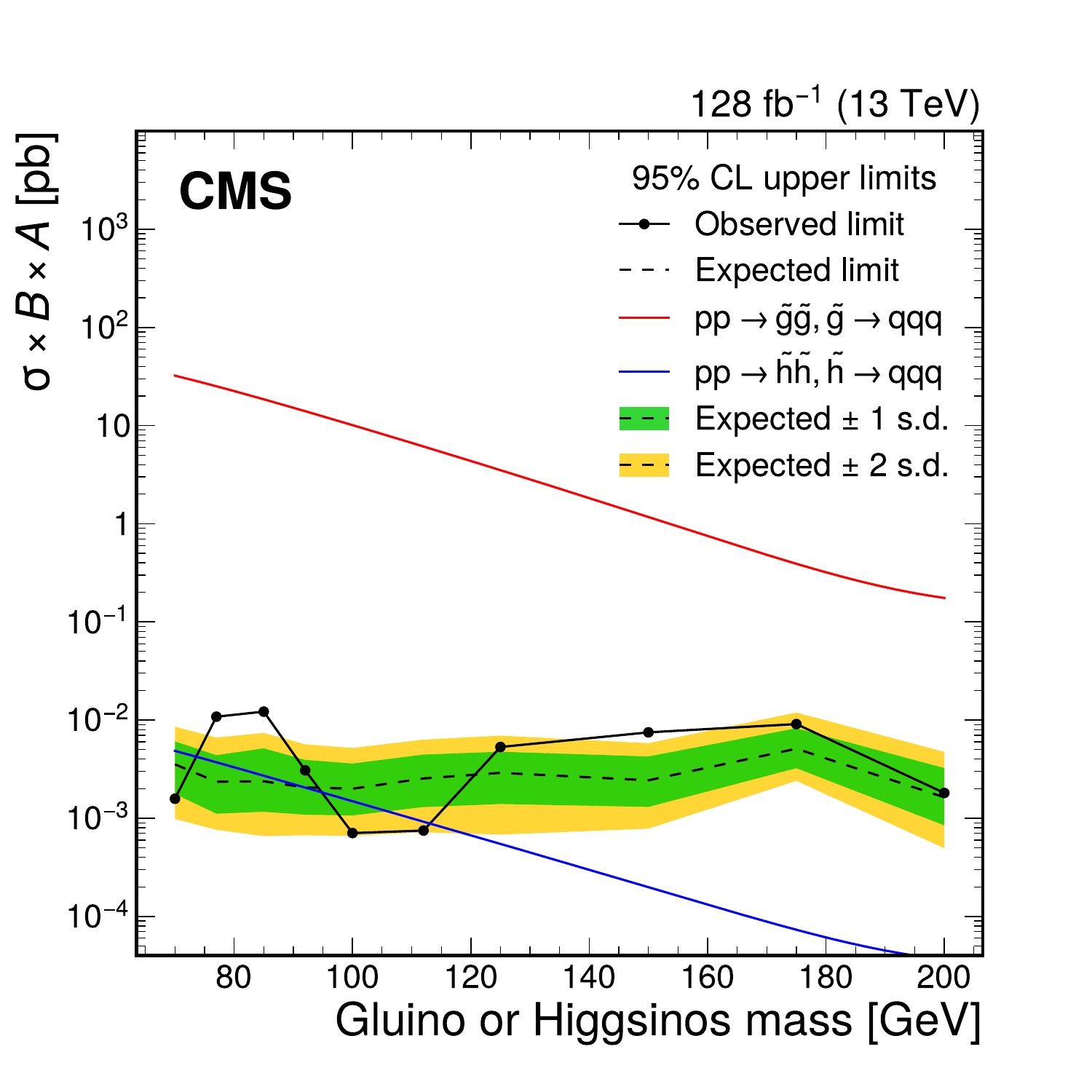}
  \includegraphics[width=0.32\textwidth]{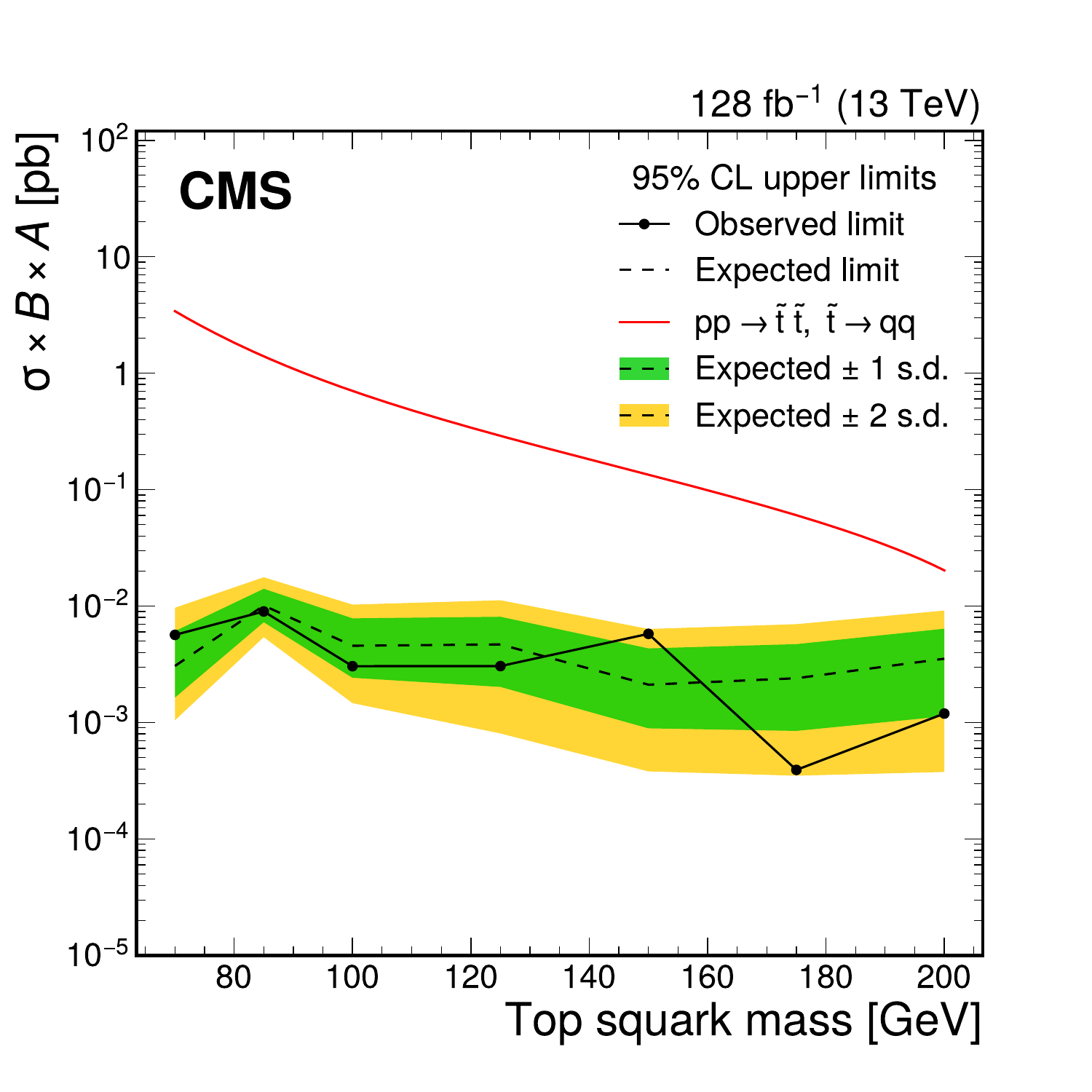}
  \includegraphics[width=0.32\textwidth]{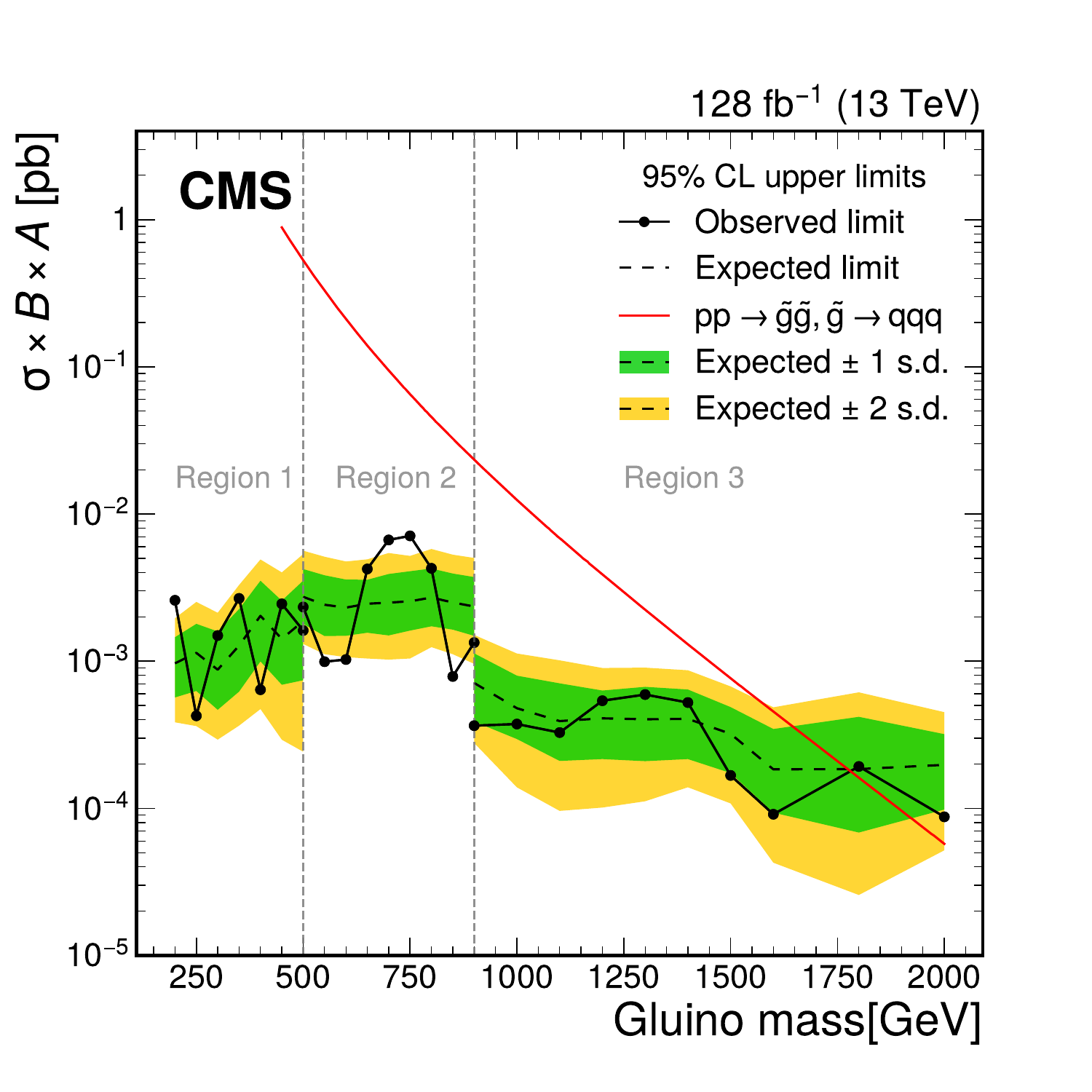}
  \vspace*{-0.20cm}
  \caption{Observed and expected 95\% confidence level upper limits on the product of the cross-section ($\sigma$), branching ratio ($B$), and acceptance ($A$) for pair-produced multijet resonances in the merged trijet (left), merged dijet (center), and resolved trijet (right) final states.}
  \label{fig:exo21004_limits}
\end{figure}

Upper limits at 95\% confidence level are set on the product of the production cross section, branching fraction, and acceptance for each of the three signatures as a function of the resonance mass, as summarized in Fig.~\ref{fig:exo21004_limits}. These limits are interpreted in $R$-parity-violating supersymmetric models with gluinos, top squarks, and higgsinos. The merged trijet search yields the first constraints on promptly decaying mass-degenerate higgsinos, excluding masses in the ranges $70$--$75$~GeV and $95$--$112$~GeV. The merged dijet channel excludes $R$-parity-violating top squarks with masses between $70$ and $200$~GeV, strengthening and extending earlier limits. Combining the merged and resolved trijet results, gluinos decaying to three quarks are excluded between $70$~GeV and $1.7$~TeV, significantly improving upon previous multijet resonance searches.

\subsection{Search for dijet resonances with Calo-scouting}
\label{sec:dijet_caloscouting}

A recent CMS analysis exploits the calorimeter-based data-scouting stream to search for narrow dijet resonances in the intermediate mass range between 0.6 and 1.8~TeV using proton--proton collision data at $\sqrt{s}=13$~TeV collected in 2016--2018, corresponding to an integrated luminosity of 117~$\mathrm{fb}^{-1}$~\cite{EXO-23-004_dijetScouting_2025}. In this stream, events are reconstructed online using calorimeter jets (Calo-jets) in the HLT and recorded in a compact format that retains only the information needed for the analysis. This minimal event content reduces the event size by about two orders of magnitude, allowing the scouting triggers to operate with a much lower $H_{\mathrm{T}}$ threshold ($H_{\mathrm{T}}>250$~GeV) and at much higher rates than the normal stream. As a result, the analysis can probe dijets in a mass window that is inaccessible for conventional high-threshold dijet searches.

\begin{figure}[!htbp]
  \centering
  \includegraphics[width=0.6\textwidth]{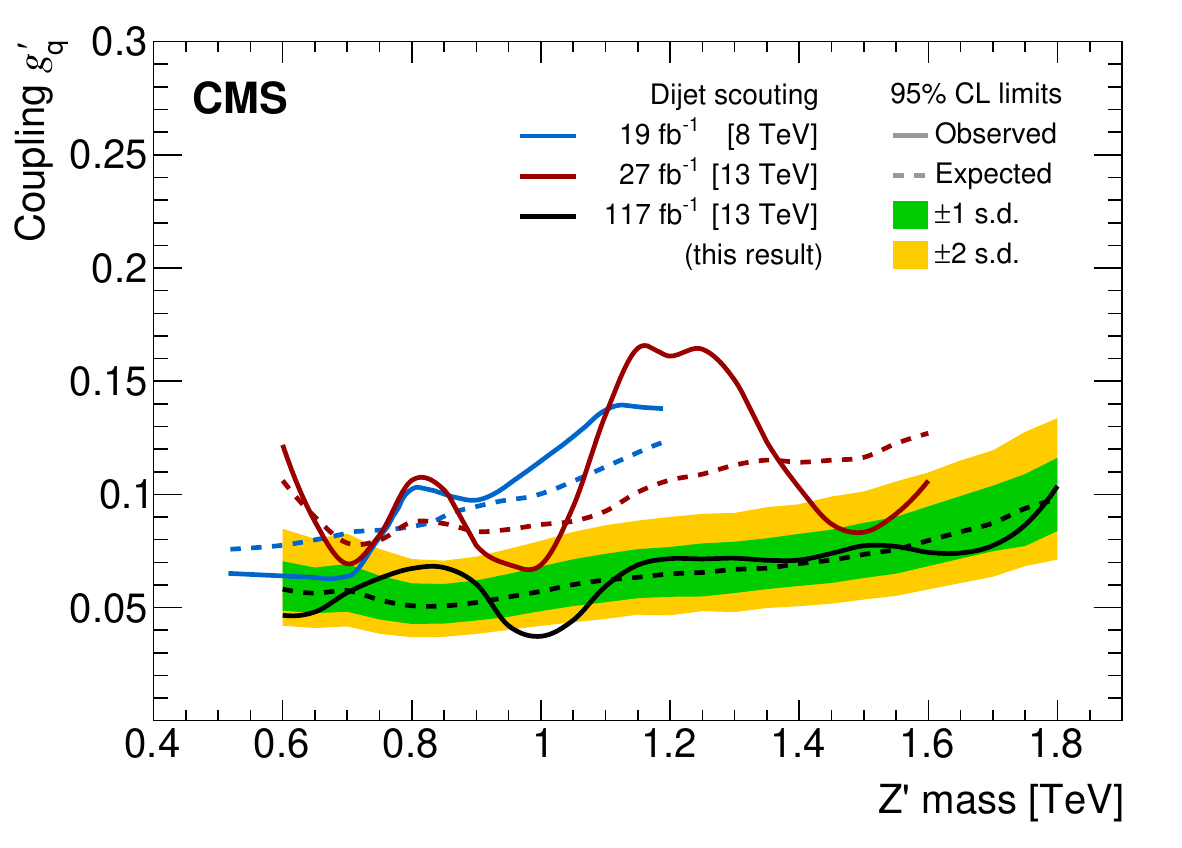}
  \vspace*{-0.25cm}
  \caption{Observed and expected 95\% confidence level upper limits on the universal quark coupling $g_q'$ as a function of the mass of a leptophobic $Z'$ boson that only couples to quarks.}
  \label{fig:exo23004_gq}
\end{figure}

The analysis sets 95\% confidence level upper limits on the product $\sigma$BA for narrow resonances decaying to quark-quark, quark-gluon, and gluon-gluon final states and excludes a broad set of benchmark models, including axigluons, colorons, scalar diquarks, excited quarks, color-octet scalars, and leptophobic $W'$ and $Z'$ bosons. In a simplified dark-matter scenario with an $s$-channel leptophobic vector mediator, the results translate into the most stringent collider limits to date on the universal quark coupling $g_\mathrm{q}'$ in this mass window. As shown in Fig.~\ref{fig:exo23004_gq}, the new limits significantly improve over earlier CMS dijet-scouting searches, reaching sensitivity to $g_\mathrm{q}'$ values as small as $\sim$0.04. This highlights the power of scouting to extend the physics reach of CMS without relying on boosted topologies.

\subsection{Search for long-lived heavy neutrinos with the B-parking data set}
\label{sec:HNL_B-Parking}

An illustrative example of the physics reach enabled by dedicated data streams is the CMS search for long-lived heavy neutrinos produced in leptonic and semileptonic decays of $B$ mesons using the 2018 ``B-parking'' dataset~\cite{EXO-22-019_HNL_Bparking_2024}. This special sample was collected with HLT paths featuring low $p_\mathrm{T}$ thresholds to enhance decays of long-lived $b$ hadrons. Events satisfying these triggers were buffered and ``parked'' for deferred offline reconstruction, yielding approximately $10^{10}$ $b\bar b$ events, well beyond what could be processed in the normal stream.

\begin{figure}[!htbp]
  \centering
  \includegraphics[width=0.44\textwidth]{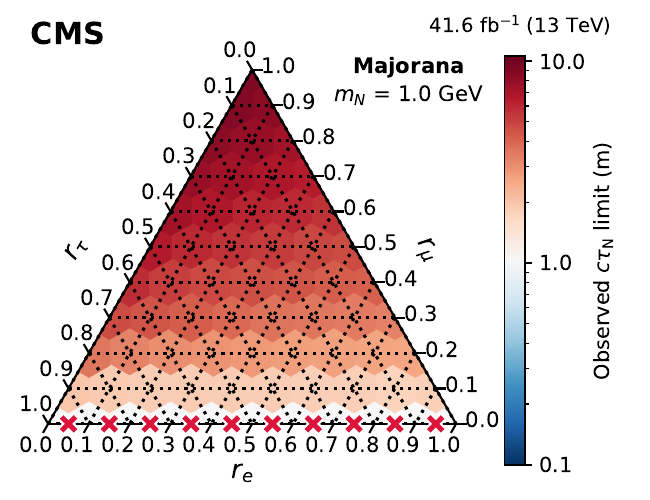}
  \includegraphics[width=0.44\textwidth]{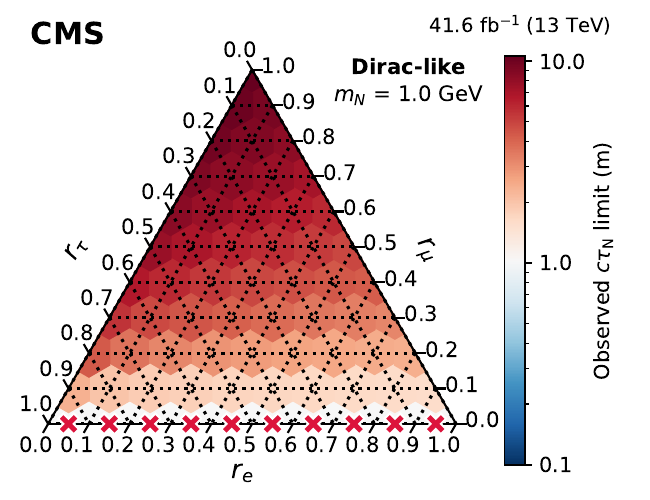}
  \vspace*{-0.15cm}
  \caption{Observed 95\% confidence level lower limits on $c\tau_\mathrm{N}$ as a function of the mixing ratios $(r_{e}, r_{\mu}, r_{\tau})$ for fixed N masses of 1GeV. The left (right) panel corresponds to the Majorana (Dirac-like) hypothesis.}
  \label{fig:exo22019_ctau}
\end{figure}

The analysis searches for heavy neutral leptons produced in $B$-meson decays that subsequently decay to a charged lepton and a pion, which results in a displaced secondary vertex inside the silicon tracker. The results are expressed as 95\% confidence level upper limits on the overall heavy-neutrino mixing strength and corresponding lower limits on the proper decay length $c\tau_{N}$, as functions of the heavy-neutrino mass and its flavour-mixing pattern. For $m_{N}=1.0$~GeV, these limits provide the most stringent collider constraints to date in scenarios where the heavy neutrino mixes predominantly with muon neutrinos, as illustrated in Fig.~\ref{fig:exo22019_ctau}.

\subsection{Search for a low-mass resonance decaying to two \texorpdfstring{$\tau$}{tau} leptons}
\label{sec:low_mass_ditau}

An inclusive CMS search is performed for a new low-mass resonance decaying to a pair of $\tau$ leptons, probing the mass range from 20 to 60~GeV for the first time at a hadron collider~\cite{CMS-PAS-EXO-24-012}. The analysis uses proton--proton collision data collected in 2022--2023 at $\sqrt{s}=13.6$~TeV. Events are selected in the Run--3 PF-scouting stream, allowing high-rate data taking with substantially lower trigger thresholds than the normal stream. Within this dataset, the search targets final states with one muonic and one hadronically decaying $\tau$ lepton, thereby maximizing acceptance for low-$p_{\mathrm{T}}$ $\tau$ leptons that would otherwise fail conventional triggers.

A key ingredient of the analysis is the introduction of the ``scouting-HPS'' algorithm (Hadron-Plus-Strip), a specialized $\tau$-reconstruction technique adapted to the compact PF-scouting event format. This algorithm reconstructs hadronic $\tau$ candidates down to transverse momenta of about 5~GeV by clustering charged and neutral hadrons and electromagnetic deposits into small-radius jets and assigning decay modes with one or three charged prongs and up to two neutral pions. By recovering such low-$p_{\mathrm{T}}$ $\tau_{\mathrm{h}}$ candidates, the scouting-HPS reconstruction extends the reach of the CMS detector to a kinematic regime that was previously inaccessible in inclusive $\tau\tau$ searches.

The results are interpreted in a benchmark model in which a scalar boson $\phi$ is produced via gluon--gluon and decays to $\tau\tau$. The analysis sets 95\% confidence level upper limits on the product $\sigma(pp\to\phi)\,B\,(\phi\to\tau\tau)$ as a function of $m_\phi$, providing the first inclusive constraints on gluon--gluon production of a low-mass scalar decaying to $\tau$-lepton pairs in this mass window. These results highlight both the effectiveness of the Run--3 PF-scouting stream and the power of dedicated reconstruction techniques in extending the CMS sensitivity to challenging low-mass signatures.

\subsection{Search for light long-lived particles decaying to displaced jets}
\label{sec:llp_displaced_jets}

A dedicated CMS search investigates light long-lived particles (LLPs) that decay to displaced jets in the inner tracker, using proton--proton collision data collected at $\sqrt{s}=13.6$~TeV~\cite{EXO-23-013_DisplacedJets_2025}. The benchmark signal is an exotic decay of the 125~GeV Higgs boson into two neutral scalars, $H\to SS$, with each $S$ decaying to a pair of quarks or tau leptons ($S\to bb$, $dd$, or $\tau\tau$). The analysis targets scalar masses between 15 and 55~GeV and mean proper decay lengths between 1~mm and 1~m, a region where previous displaced-jet searches had limited sensitivity. Events are collected with a set of dedicated displaced-jet triggers that require large hadronic activity and one or more jets with tracks inconsistent with the primary vertex, including a trigger path seeded by a low-$p_{\mathrm{T}}$ muon to enhance the acceptance for heavy-flavor decays.

\begin{figure}[!htbp]
  \centering
  \includegraphics[width=0.32\textwidth]{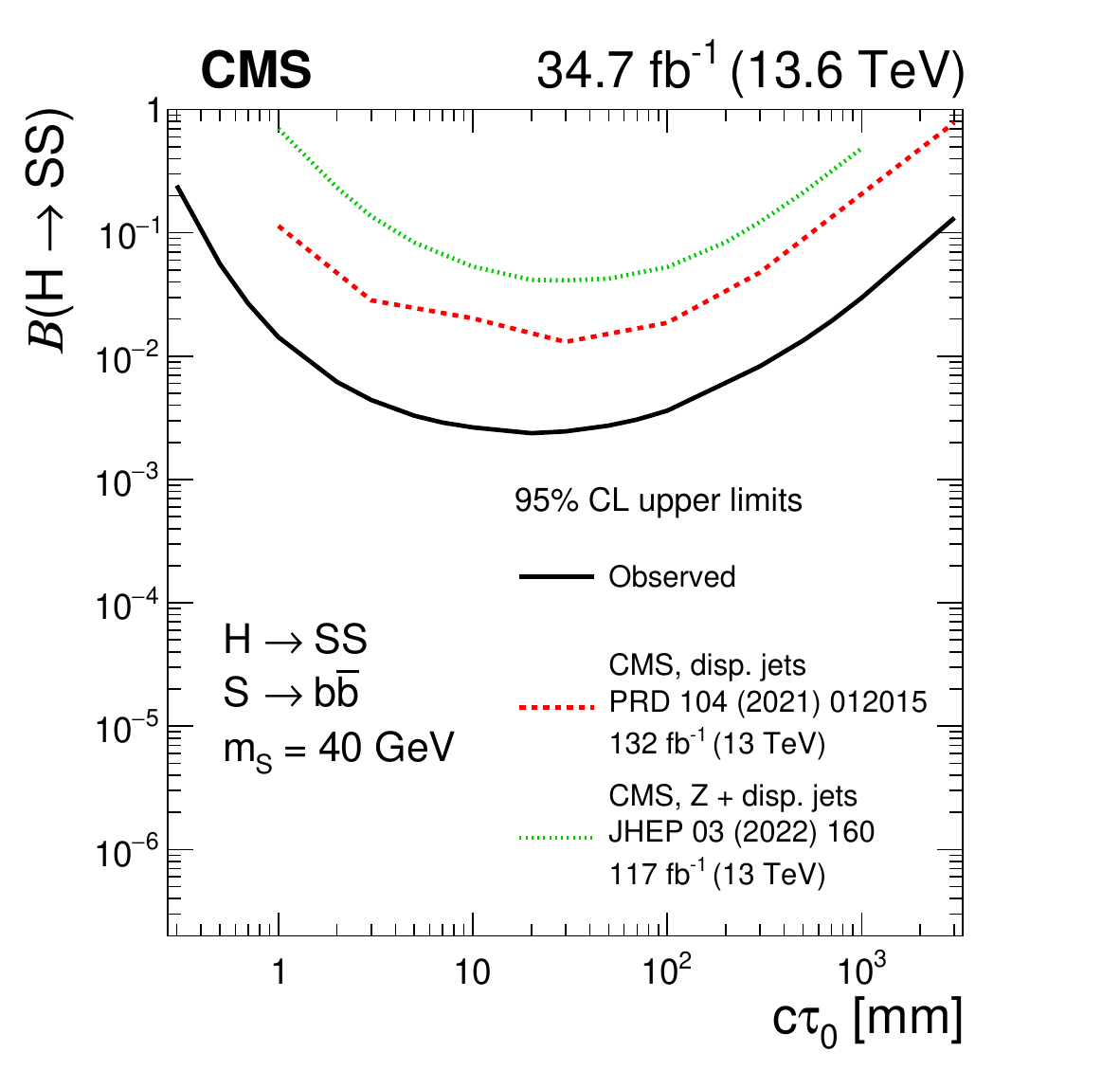}
  \includegraphics[width=0.32\textwidth]{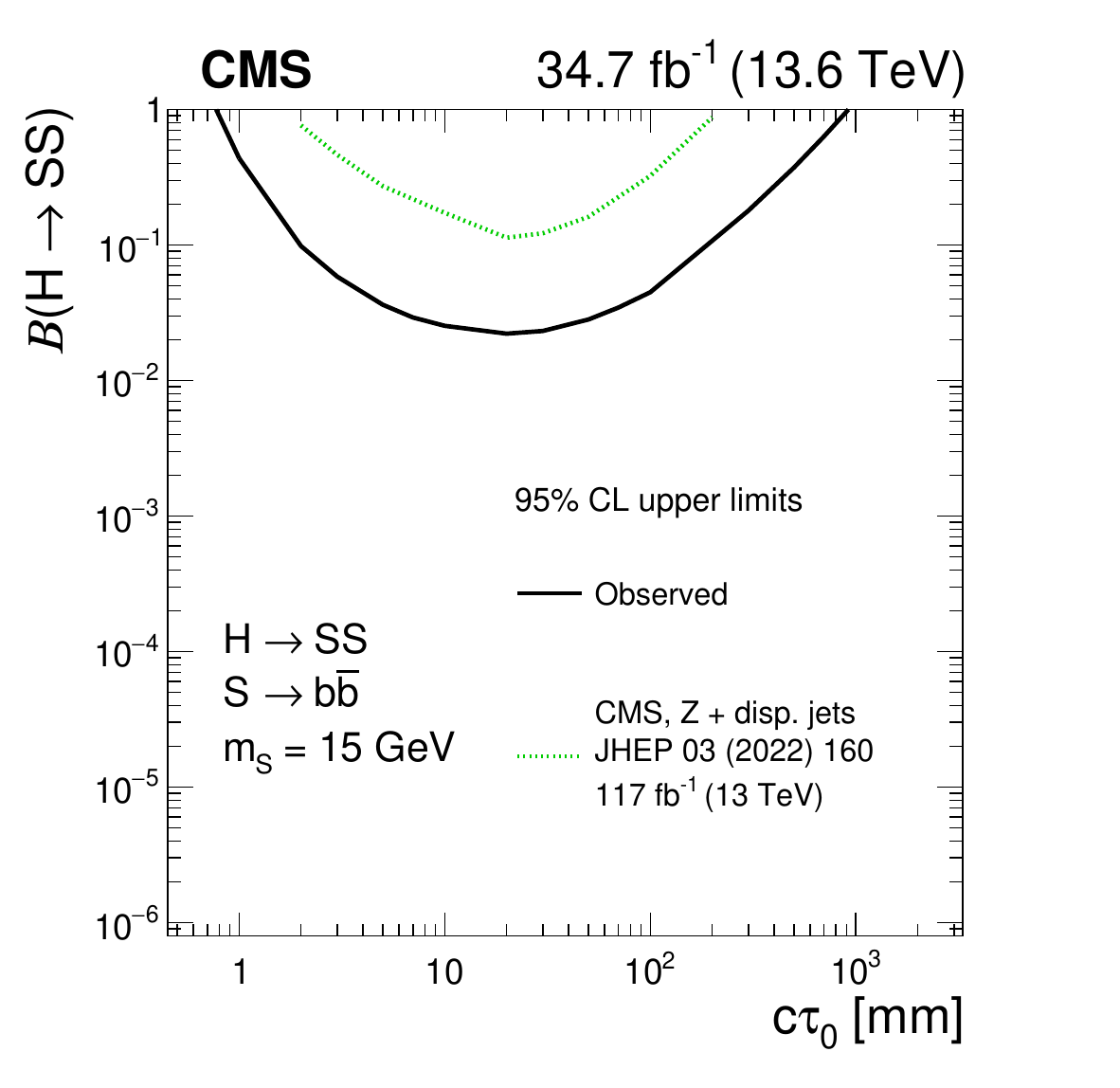}
  \includegraphics[width=0.32\textwidth]{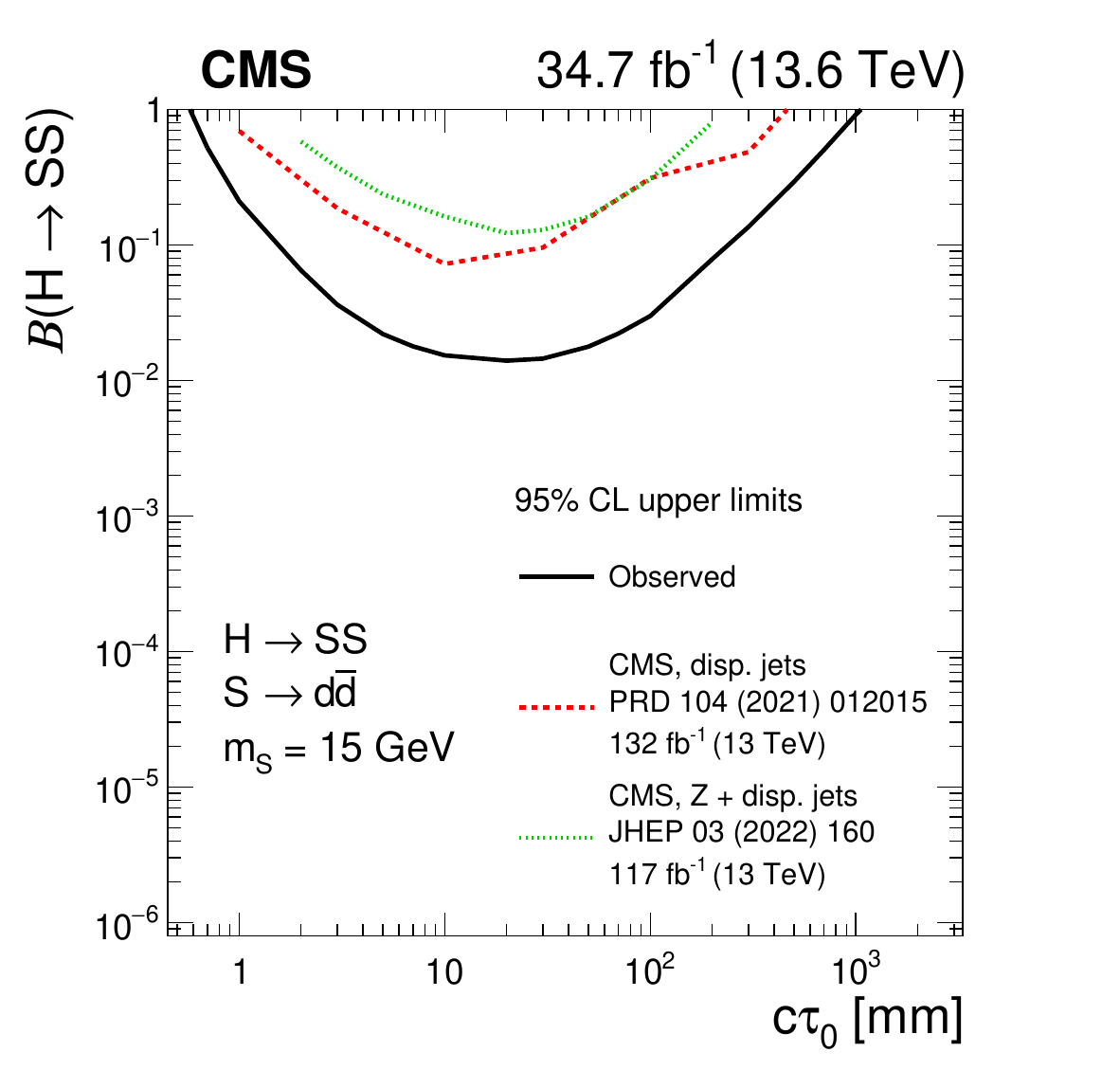}
  \caption{Observed 95\% confidence level upper limits on the branching fraction $B(H\to SS)$ as a function of the LLP proper decay length $c\tau_{0}$ for representative signal points. The three panels show the cases $S\to bb$ with $m_{S}=40$~GeV (left), $S\to bb$ with $m_{S}=15$~GeV (center), and $S\to dd$ with $m_{S}=15$~GeV (right). }
  \label{fig:exo23013_limits}
\end{figure}

The analysis sets 95\% confidence level upper limits on the branching fraction $\mathcal{B}(H\to SS)$ as a function of the LLP mass and lifetime, separately for the $S\to bb$, $S\to dd$, and $S\to\tau\tau$ decay modes. For $S\to bb$ and $S\to dd$, the search reaches its best sensitivity for $m_{S}\gtrsim 20$~GeV and decay lengths of a few centimeters, excluding branching fractions of order 1\% or lower across a broad range of lifetimes. As illustrated in Fig.~\ref{fig:exo23013_limits}, the new limits improve upon previous CMS displaced-jet and $Z$+displaced-jet searches. The analysis also provides the first constraints on hadronically decaying displaced tau leptons with decay lengths below about 1~m.

\section{Summary}
\label{sec:summary}

Dedicated data streams have become an essential component of the CMS physics program at the LHC. By relaxing the tight coupling between trigger thresholds, event size, and prompt reconstruction, the data-scouting and data-parking strategies allow the experiment to record very large samples in challenging regions of phase space that are difficult to access with the normal data stream. The scouting streams operate at much higher rates by storing a compact HLT representation of the events, while the parking streams preserve the full detector readout for deferred offline reconstruction. Together with the planned Level-1 scouting system for the Phase-2 upgrade, these techniques provide a flexible framework that can be tailored to a wide variety of signatures beyond the Standard Model.

The physics reach of these approaches is illustrated by several representative CMS searches. Data-scouting streams push down trigger thresholds and enable stringent constraints on low-mass dijet and multijet resonances, as well as the first inclusive search for a low-mass scalar decaying to $\tau$-lepton pairs. The 2018 B-parking campaign, based on parked events collected with dedicated low-$p_{\mathrm{T}}$ triggers, yields an unprecedented sample of $B$-hadron decays and leads to the most stringent limits to date on long-lived heavy neutrinos produced in $B$-meson decays. In Run~3, specialized triggers and reconstruction techniques further extend this program by providing sensitivity to light long-lived particles that decay into displaced jets in the tracker. Although no significant deviations from the Standard Model have been observed so far, these analyses demonstrate that dedicated data streams substantially extend the discovery potential of CMS, and they will play an increasingly important role as the LHC moves toward the high-luminosity era.

\FloatBarrier

\printbibliography

\end{document}